\documentclass[12pt]{article}
\usepackage{graphicx}
\usepackage{here}
\newcommand{\lesssim}{\raisebox{0.3mm}{\em $\, <$} \hspace{-3.2mm}
\raisebox{-1.4mm}{\em $\sim \,$}}
\begin{document}

\begin{center}
{\large\bf
The possibility to observe the non-standard interaction
by the Hyperkamiokande atmospheric neutrino experiment
}
\end{center}

\begin{center} {
               \large{\sc
Shinya Fukasawa} and 
               \large{\sc
Osamu Yasuda} }
\end{center}
\vspace*{0cm}
{\it
\begin{center}

Department of Physics, Tokyo Metropolitan University,\\
Minami-Osawa, Hachioji, Tokyo 192-0397, Japan
\end{center}}

\vspace*{0.5cm}

{\Large \bf
\begin{center} Abstract \end{center}  }

It was suggested that a tension between the mass-squared differences
obtained from the solar neutrino and KamLAND experiments can be solved
by introducing the non-standard flavor-dependent interaction in
neutrino propagation.
In this paper we discuss the possibility to test
such a hypothesis by atmospheric neutrino observations
at the future Hyper-Kamiokande experiment.
Assuming that the mass hierarchy is known,
we find that the best-fit value from the solar neutrino and KamLAND data
can be tested at more than 8 $\sigma$, while
the one from the global analysis can be examined at
5.0 $\sigma$ (1.4 $\sigma$) for the normal (inverted)
mass hierarchy.

\vspace*{.5cm}

\newpage
\section{Introduction}
It is well established by solar, atmospheric, reactor and accelerator neutrino
experiments that neutrinos have masses and mixings\,\cite{Agashe:2014kda}.
In the standard three flavor neutrino oscillation framework,
there are three mixing angles $\theta_{12}$, $\theta_{13}$, $\theta_{23}$
and two mass-squared differences $\Delta m^2_{31}$, $\Delta m^2_{21}$.
Their approximate values are determined as
$(\Delta m^2_{21},\sin^22\theta_{12}) \simeq (7.5\times 10^{-5}$eV$^2$,$0.86)$,
$(|\Delta m^2_{31}|,\sin^22\theta_{23}) \simeq (2.5\times 10^{-3}$eV$^2$,$1.0)$,
$\sin^22\theta_{13}\simeq 0.09$.
However we do not know the value of the Dirac CP phase $\delta_{\rm CP}$,
the sign of $\Delta m^2_{31}$ (the mass hierarchy) and the octant
of $\theta_{23}$ (the sign of $\pi/4-\theta_{23}$).
Future neutrino oscillation experiments
with high statistics\,\cite{Abe:2011ts,Acciarri:2015uup}
are planned to measure these undetermined
neutrino oscillation parameters and we are entering an era of the
precision measurements.  With these precision measurements, we
can probe the new physics by looking at the deviation from
the standard three flavor neutrino mixing scenario.

Flavor-dependent neutrino NonStandard Interactions (NSI) have been
studied as the new physics candidates which may be searched at the
future neutrino experiments.  There are two types of NSI.  One is a
neutral current nonstandard interaction\,\cite{Wolfenstein:1977ue,Guzzo:1991hi,Roulet:1991sm} and the other is a charged
current nonstandard interaction\,\cite{Grossman:1995wx}.  The neutral current NSI affects the
neutrino propagation through the matter effect and hence experiments
with a long baseline such as atmospheric neutrino and LBL experiments
are expected to have the sensitivity to the neutral current NSI.  On the
other hand, the charged current NSI causes zero distance effects in
neutrino oscillation.  In this paper, we concentrate on the effects of
neutral current NSI in neutrino propagation and study the sensitivity
of the future atmospheric neutrino experiments Hyper-Kamiokande to NSI
with a parametrization introduced to study solar neutrinos.

It was pointed out in Ref.\,\cite{Gonzalez-Garcia:2013usa} that there is a
tension between the mass-squared difference deduced from the
solar neutrino observations and the one from the KamLAND experiment, and that the tension can be
resolved by introducing the flavor-dependent NSI
in neutrino propagation.
Such a hint for NSI gives us a strong motivation to study NSI in propagation in details.

In Ref.\,\cite{Fukasawa:2015jaa} it was shown that the atmospheric
neutrino measurements at Hyper-Kamiokande has a very good sensitivity
to the NSI, on the assumptions that (i) all the $\mu$ components of the NSI
vanish and (ii) the ($\tau$, $\tau$) component is expressed in terms of
the other components as is suggested by the high energy atmospheric
neutrino data.  In this paper we discuss the sensitivity of the
atmospheric neutrino measurements at Hyper-Kamiokande to NSI without
the assumptions (i) and (ii) mentioned above.  Since the parametrization which is
used in Ref.\,\cite{Gonzalez-Garcia:2013usa} is different from the
ordinary one in the three flavor basis, a non-trivial mapping
is required to compare the results in these two parametrizations.
Our analysis was performed by taking this non-trivial mapping
into account.

Constraints on $\epsilon_{\alpha\beta}$
have been discussed by many people in the past.
from atmospheric neutrinos\,\cite{GonzalezGarcia:1998hj,Lipari:1999vh,Fornengo:1999zp,Fornengo:2001pm,GonzalezGarcia:2004wg},
from $e^+ e^-$ colliders\,\cite{Berezhiani:2001rs},
from the compilation of various neutrino data\,\cite{Davidson:2003ha},
from solar neutrinos\,\cite{Friedland:2004pp,Miranda:2004nb,Palazzo:2009rb},
from $\nu_e e$ or $\bar{\nu}_e e$ scatterings\,\cite{Barranco:2005ps,Barranco:2007ej},
from solar and reactor neutrinos\,\cite{Bolanos:2008km},
from solar, reactor and accelerator neutrinos\,\cite{Escrihuela:2009up}.
The constraints on $\epsilon_{ee}$ and $\epsilon_{e\tau}$
from the atmospheric neutrino has been discussed in
Ref.~\cite{GonzalezGarcia:2011my} along with those
from the long-baseline experiments,
in Ref.~\cite{Mitsuka:2011ty} by the Super-Kamiokande Collaboration,
in Ref.~\cite{Ohlsson:2013epa,Esmaili:2013fva,Choubey:2014iia}
on the future extension of the IceCube experiment,
in Ref.~\cite{Chatterjee:2014gxa} on the future experiment
with the iron calorimeter or liquid argon detectors,
with the ansatz different from ours.

This paper is organized as follows. In Section \ref{3nu_nsi}, we
describe the current knowledge and constraints on NSI in propagation
from solar neutrinos and atmospheric neutrinos. In Section
\ref{analysis}, we study the sensitivity of the future atmospheric
neutrino experiment Hyper-Kamiokande to NSI. In Section
\ref{conclusion}, we draw our conclusions.  
In the appendix\,\ref{appendixa}, we derive the relation
between the two different parametrizations of NSI.

\section{Three flavor neutrino oscillation framework with NSI\label{3nu_nsi}}
\subsection{Nonstandard interactions\label{nsi_intro}}
Let us start with the effective flavor-dependent neutral current
neutrino nonstandard interactions in propagation given by
\begin{eqnarray}
{\cal L}_{\mbox{\rm\scriptsize eff}}^{\mbox{\tiny{\rm NSI}}} 
=-2\sqrt{2}\, \epsilon_{\alpha\beta}^{ff'P} G_F
\left(\overline{\nu}_{\alpha L} \gamma_\mu \nu_{\beta L}\right)\,
\left(\overline{f}_P \gamma^\mu f_P'\right),
\label{NSIop}
\end{eqnarray}
where $f_P$ and $f_P'$ stand for fermions with chirality $P$ and
$\epsilon_{\alpha\beta}^{ff'P}$ is a dimensionless constant
which is normalized by the Fermi coupling constant $G_F$.
The presence of NSI (\ref{NSIop}) modifies the MSW potential in the
flavor basis:
\begin{eqnarray}
 \sqrt{2} G_F N_e\left(
\begin{array}{ccc}
1 &  0& 0\\
0 &0  &0\\
0 &0 & 0
\end{array}
\right)
\rightarrow
{\cal A}\,,
\end{eqnarray}
where
\begin{eqnarray}
{\cal A} \equiv
\sqrt{2} G_F N_e \left(
\begin{array}{ccc}
1+ \epsilon_{ee} & \epsilon_{e\mu} & \epsilon_{e\tau}\\
\epsilon_{\mu e} & \epsilon_{\mu\mu} & \epsilon_{\mu\tau}\\
\epsilon_{\tau e} & \epsilon_{\tau\mu} & \epsilon_{\tau\tau}
\end{array}
\right),
\label{matter-np}
\end{eqnarray}
$\epsilon_{\alpha\beta}$ is defined by
\begin{equation}
\epsilon_{\alpha\beta}\equiv\sum_{f=e,u,d}\frac{N_f}{N_e}\epsilon_{\alpha\beta}^{f}\,,
\end{equation}
and $N_f~(f=e, u, d)$ stands for number densities of fermions $f$.
Here we defined the new NSI parameters as
$\epsilon_{\alpha\beta}^{fP}\equiv\epsilon_{\alpha\beta}^{ffP}$ and
$\epsilon_{\alpha\beta}^{f}\equiv\epsilon_{\alpha\beta}^{fL}+\epsilon_{\alpha\beta}^{fR}$
since the matter effect is sensitive only to the coherent scattering
and only to the vector part in the interaction.  As can be seen from
the definition of $\epsilon_{\alpha\beta}$, the neutrino oscillation
experiments on the Earth are sensitive only to the sum of
$\epsilon_{\alpha\beta}^{f}$.  We call the most general
parametrization (\ref{matter-np}) of NSI in the flavor basis
the standard NSI parametrization in this paper. In the three flavor
neutrino oscillation framework with NSI, the neutrino evolution is
governed by the Dirac equation:
\begin{eqnarray}
i {d \over dx} \left( \begin{array}{c} \nu_e(x) \\ \nu_{\mu}(x) \\ 
\nu_{\tau}(x)
\end{array} \right) = 
\left[  U {\rm diag} \left(0, \Delta E_{21}, \Delta E_{31}
\right)  U^{-1}
+{\cal A}\right]
\left( \begin{array}{c}
\nu_e(x) \\ \nu_{\mu}(x) \\ \nu_{\tau}(x)
\end{array} \right)\,,
\label{eqn:sch}
\end{eqnarray}
where
$U$ is the leptonic mixing matrix defined by
\begin{eqnarray}
U&\equiv&\left(
\begin{array}{ccc}
c_{12}c_{13} & s_{12}c_{13} &  s_{13}e^{-i\delta_{\rm CP}}\cr
-s_{12}c_{23}-c_{12}s_{23}s_{13}e^{i\delta_{\rm CP}} & 
c_{12}c_{23}-s_{12}s_{23}s_{13}e^{i\delta_{\rm CP}} & s_{23}c_{13}\cr
s_{12}s_{23}-c_{12}c_{23}s_{13}e^{i\delta_{\rm CP}} & 
-c_{12}s_{23}-s_{12}c_{23}s_{13}e^{i\delta_{\rm CP}} & c_{23}c_{13}
\end{array}
\right),
\label{eqn:mns3}
\end{eqnarray}
and $\Delta E_{jk}\equiv\Delta m_{jk}^2/2E\equiv (m_j^2-m_k^2)/2E$,
$c_{jk}\equiv\cos\theta_{jk}$, $s_{jk}\equiv\sin\theta_{jk}$.

\subsection{Solar neutrinos\label{sol_nu}}
In Refs.\,\cite{Gonzalez-Garcia:2013usa,Maltoni:2015kca}
it was pointed out that there is a
tension between the two mass squared differences
extracted from the KamLAND and solar neutrino experiments.
The mass squared difference $\Delta m^2_{21}$
($=4.7\times10^{-3} {\rm eV}^2$) extracted from the
solar neutrino data 
is $2\sigma$ smaller than that
from the KamLAND data $\Delta m^2_{21}$ ($=7.5\times10^{-3} {\rm eV}^2$).
The authors of Refs.\,\cite{Gonzalez-Garcia:2013usa,Maltoni:2015kca}
discussed the tension can be removed by introducing NSI in propagation.

To discuss the effect of NSI on solar neutrinos, we
reduce the $3 \times 3$ Hamiltonian in the Dirac equation
Eq.\,(\ref{eqn:sch}) to an effective $2 \times 2$ Hamiltonian to get
the survival probability $P(\nu_e \rightarrow \nu_e)$ because solar
neutrinos are approximately driven by one mass squared difference
$\Delta m_{21}^2$\,\cite{Gonzalez-Garcia:2013usa}.  The survival probability $P(\nu_e \rightarrow
\nu_e)$ can be written as
\begin{eqnarray}
P(\nu_e \rightarrow \nu_e) = c_{13}^4 P_{\rm eff} +  s_{13}^4.
\end{eqnarray}
$P_{\rm eff}$ can be calculated by using the effective $2 \times 2$ Hamiltonian $H^{\rm eff}$ written as
\begin{eqnarray*}
H^{\rm eff}=
\frac{\Delta m^2_{21}}{4E}\left(\begin{array}{cc}
-\cos2\theta_{12} & \sin2\theta_{12}  \\
\sin2\theta_{12} & \cos2\theta_{12}
\end{array}\right) 
+
\left(\begin{array}{cc}
c^2_{13} A & 0 \\
0 & 0
\end{array}\right)  + 
 A\sum_{f=e,u,d} \frac{N_f}{N_e}
\left(\begin{array}{cc}
- \epsilon_{D}^f &  \epsilon_{N}^f \\
 \epsilon_{N}^{f*} &  \epsilon_{D}^f
\end{array}\right),
\end{eqnarray*}
where  $\epsilon^f_{D}$ and $\epsilon^f_{N}$ are linear combinations of the standard NSI parameters:
\begin{eqnarray}
\epsilon_{D}^f &=&c_{13}s_{13}{\rm Re}\left[ e^{i\delta_{\rm CP}}\left(s_{23}\epsilon_{e \mu}^f+c_{23}\epsilon_{e \tau}^f\right) \right]-\left(1+s_{13}^2\right)c_{23}s_{23}{\rm Re}\left[\epsilon_{\mu \tau}^f\right]  \nonumber \\
& &-\frac{c_{13}^2}{2}\left(\epsilon_{e e}^f-\epsilon_{\mu \mu}^f\right)+\frac{s_{23}^2-s_{13}^2c_{23}^2}{2}\left(\epsilon_{\tau \tau}^f-\epsilon_{\mu \mu}^f\right) \nonumber \\
\epsilon_{N}^f&=& c_{13}\left(c_{23}\epsilon_{e \mu}^f-s_{23}\epsilon_{e\tau}^f \right)+s_{13}e^{-i\delta_{\rm CP}}\left[ s_{23}^2\epsilon_{\mu\tau}^f-c_{23}^2\epsilon_{\mu\tau}^{f*} +c_{23}s_{23}\left(\epsilon_{\tau \tau}^f-\epsilon_{\mu \mu}^f\right) \right].
\label{epsilonn}
\end{eqnarray}
Ref.\,\cite{Gonzalez-Garcia:2013usa,Maltoni:2015kca} discussed the sensitivity of solar neutrino and KamLAND experiments to $\epsilon_{D}^f$ and real $\epsilon_{N}^f$ for one particular choice of $f=u$ or $f=d$ at a time.
The best fit values from the solar neutrino and KamLAND data are $(\epsilon_{D}^u,\epsilon_{N}^u)=(-0.22,-0.30)$ and $(\epsilon_{D}^d,\epsilon_{N}^d)=(-0.12,-0.16)$ and that from the global analysis of the neutrino oscillation data are $(\epsilon_{D}^u,\epsilon_{N}^u)=(-0.140, -0.030)$ and $(\epsilon_{D}^d,\epsilon_{N}^d)=(-0.145, -0.036)$.
These results give us a hint for the existence of NSI.
In addition to the above, Ref.\,\cite{Gonzalez-Garcia:2013usa,Maltoni:2015kca} also discussed the possibility of the dark-side solution ($\Delta m^2_{21}<0$ and $\theta_{21}>\pi/4$) which requires NSI in the solar neutrino problem.
The allowed regions for the dark-side solution are disconnected from that for the standard LMA solution in the plane $(\epsilon_{D}^f,\epsilon_{N}^f)$ and those for the dark-side solution within $3\sigma$ do not contain the standard scenario $\epsilon_{D}^f=\epsilon_{N}^f=0$.

\subsection{Atmospheric neutrinos\label{atm_nu}}
In this subsection, we describe the constraints on NSI from the atmospheric neutrino experiments and introduce a relation between $\epsilon_{ee}$,  $|\epsilon_{e\tau}|$ and  $\epsilon_{\tau\tau}$ and
 a matter angle $\beta$.
 Atmospheric neutrinos go through the Earth and interact with electrons, up and down quarks.
In the Earth, the number densities of electrons, protons and neutrons are approximately equal and hence those of up quarks and down quarks are approximately the same.
From these, one can define $\epsilon_{\alpha\beta}$ as
\begin{eqnarray}
\label{eps_atm}
\epsilon_{\alpha\beta}=\epsilon_{\alpha\beta}^{e}+3\epsilon_{\alpha\beta}^{u}+3\epsilon_{\alpha\beta}^{d},
\end{eqnarray}
and we have the following limits \cite{Biggio:2009nt} on $\epsilon_{\alpha\beta}$ at 90\% C.L.:
\begin{eqnarray}
\left(
\begin{array}{lll}
|\epsilon_{ee}| < 4\times 10^0 & \quad|\epsilon_{e\mu}| < 3\times 10^{-1}
& \quad|\epsilon_{e\tau}| < 3\times 10^{0\ }\\
& \quad |\epsilon_{\mu\mu}| < 7\times 10^{-2}
& \quad|\epsilon_{\mu\tau}| < 3\times 10^{-1}\\
& & \quad|\epsilon_{\tau\tau}| < 2\times 10^{1\ }
\end{array}
\right).
\label{epsilon-m}
\end{eqnarray}

To investigate the sensitivity of the atmospheric neutrino experiment to $\epsilon^f_{D}$ and $\epsilon^f_{N}$, we have to convert $\epsilon^f_{D}$ and $\epsilon^f_{N}$ into $\epsilon^f_{\alpha\beta}$ because $\epsilon^f_{D}$ and $\epsilon^f_{N}$ are valid only in the solar neutrinos analysis.
$\epsilon^f_{D}$ and $\epsilon^f_{N}$ are expressed in terms of $\epsilon^f_{\alpha\beta}$ as the following:
\begin{eqnarray}
&{\ }&
\hspace*{-10mm}
|\epsilon_{e\tau}^f|=\frac{\sin\left(\phi^f_{\mu\tau}\right) t_{13}}{\sin\left(\delta_{cp}+\phi^f_{e\tau}\right) s_{23}}|\epsilon_{\mu\tau}^f| 
+\frac{\sin\left(\delta_{cp}+\phi^f_{e\mu}\right)}{ t_{23}\sin\left(\delta_{cp}+\phi^f_{e\tau}\right)}|\epsilon_{e\mu}^f| 
\nonumber \\  
&{\ }&
\hspace*{3mm}
-\frac{\sin\left(\delta_{cp}+\psi^f\right)}
{\sin\left(\delta_{cp}+\phi^f_{e\tau}\right) s_{23} c_{13}}|\epsilon_{N}^f|,  
\nonumber \\  
&{\ }&
\hspace*{-10mm}
\epsilon_{\tau\tau}^f - \epsilon_{\mu\mu}^f
= 2\left\{ \frac{\cos\phi^f_{\mu\tau}}{\tan2\theta_{23}}
+\frac{\sin\phi^f_{\mu\tau}}{\tan\left(\delta_{cp}+\phi^f_{e\tau}\right)
\sin2\theta_{23}} \right\}|\epsilon_{\mu\tau}^f|  
\nonumber \\  
&{\ }&
\hspace*{13mm}
+\frac{\sin\left(\phi^f_{e\mu}-\phi^f_{e\tau}\right)}
{s_{23} t_{13}\sin\left(\delta_{cp}+\phi^f_{e\tau}\right)}|\epsilon_{e\mu}^f|
- \frac{2\sin(\psi^f-\phi^f_{e\tau})}{\sin(\delta_{cp}+\phi^f_{e\tau})s_{13}\sin2\theta_{23}}|\epsilon_{N}^f|, 
\nonumber \\ 
&{\ }&
\hspace*{-10mm}
\epsilon_{e e}^f-\epsilon_{\mu \mu}^f=
2\left[
\frac{s_{23}^2-s_{13}^2c_{23}^2}{c^2_{13}}
\left\{
\frac{\cos\phi^f_{\mu\tau}}{\tan2\theta_{23}}
+\frac{\sin\phi^f_{\mu\tau}}{\tan\left(\delta_{cp}+\phi^f_{e\tau}\right)
\sin2\theta_{23}}
\right\}\right.
\nonumber\\
&{\ }&
\hspace*{15mm}
\left.+\frac{t_{13}^2}{t_{23}}\cdot
\frac{\sin\phi^f_{\mu\tau}}{\tan\left(\delta_{cp}+\phi^f_{e\tau}\right)}
-\frac{1+s_{13}^2}{2c_{13}^2}\sin2\theta_{23}\cos\phi_{\mu\tau}
\right]|\epsilon_{\mu\tau}^f|
\nonumber\\
&{\ }&
\hspace*{10mm}
+2\left[
\frac{s_{23}^2-s_{13}^2c_{23}^2}{s_{23}\sin2\theta_{13}}\cdot
\frac{\sin\left(\phi^f_{e\mu}-\phi^f_{e\tau}\right)}
{\sin\left(\delta_{cp}+\phi^f_{e\tau}\right)}
+t_{13}s_{23}\cos\left(\delta_{cp}+\phi^f_{e\mu}\right)
\right.
\nonumber\\
&{\ }&
\hspace*{15mm}
\left.+\frac{t_{13}c_{23}}{t_{23}}\cdot
\frac{\sin(\delta_{cp}+\phi^f_{e\mu})}
{\tan\left(\delta_{cp}+\phi^f_{e\tau}\right)
\sin2\theta_{23}}
\right]|\epsilon_{e\mu}^f|
\nonumber\\
&{\ }&
\hspace*{10mm}
-2\left[
\frac{s_{23}^2-s_{13}^2c_{23}^2}{s_{13}c_{13}^2\sin2\theta_{23}}\cdot
\frac{\sin\left(\psi^f-\phi^f_{e\tau}\right)}
{\sin\left(\delta_{cp}+\phi^f_{e\tau}\right)}\right.
\nonumber\\
&{\ }&
\hspace*{15mm}
\left.+\frac{t_{13}}{t_{23}c_{13}}\cdot
\frac{\sin(\delta_{cp}+\psi^f)}{\tan\left(\delta_{cp}+\phi^f_{e\tau}\right)}
\right]|\epsilon_N^f|
-\frac{2}{c_{13}^2}\epsilon_{D}^f\,,
\label{eq_nsi1}
\end{eqnarray}
where $\psi^f={\rm arg} (\epsilon_{N}^f)$, $\phi^f_{\alpha\beta}={\rm arg} (\epsilon_{\alpha\beta}^f)$ and $t_{ij}\equiv\tan\theta_{ij}$.
When we consider only one particular choice of $f=u$ or $f=d$ at a time as in Ref.\,\cite{Gonzalez-Garcia:2013usa}, from the definition of $\epsilon_{\alpha\beta}$ (\ref{eps_atm}), we cannot distinguish the case of $f=u$ from that of $f=d$ in the atmospheric neutrinos analysis.
Therefore we concentrate on only one particular choice of $f=d$ in this paper and then we have
\begin{eqnarray}
&\epsilon_{\alpha\beta}&=3\epsilon_{\alpha\beta}^{d}\nonumber \\
&\phi_{\alpha\beta}& \equiv {\rm arg}\left(\epsilon_{\alpha\beta}\right)={\rm arg}\left(\epsilon_{\alpha\beta}^d\right)\nonumber \\
&\epsilon_{D}&=\epsilon_{D}^{d}\nonumber \\
&\epsilon_{N}&=\epsilon_{N}^{d}\nonumber \\
&\psi& \equiv {\rm arg}\left(\epsilon_{N}^d\right).
\label{eq_nsi2}
\end{eqnarray}

\subsubsection{The case with $\epsilon_{\alpha\mu}=0~(\alpha=e,\mu,\tau)$}
It was pointed out in Refs.\,\cite{Friedland:2004ah,Friedland:2005vy}
that if the $\mu$ components of $\epsilon_{\alpha\beta}$ are set to zero then the high-energy atmospheric neutrino data, where the matter effects are dominant, are consistent with NSI only when the following inequality is hold:
\begin{equation}
\label{eq:ansatz_b}
\min_{\pm}
\left(
\left|1+\epsilon_{ee}+\epsilon_{\tau\tau}\pm\sqrt{(1+\epsilon_{ee}-\epsilon_{\tau\tau})^2+4|\epsilon_{e\tau}|^2}\right|
\right) \lesssim 0.4\,,
\end{equation}
where the arguments of the absolute value on the left hand side
are the two nonzero eigenvalues of the matrix ${\cal A}$
in the absence of $\epsilon_{\alpha\mu}~(\alpha=e,\mu,\tau)$
component, and the $+$ ($-$) sign in $\pm$ is chosen
when $1+\epsilon_{ee}+\epsilon_{\tau\tau}$ is negative (positive).
Notice that in the limit of
\begin{equation}
\min_{\pm}
\left(
\left|1+\epsilon_{ee}+\epsilon_{\tau\tau}\pm\sqrt{(1+\epsilon_{ee}-\epsilon_{\tau\tau})^2+4|\epsilon_{e\tau}|^2}\right|
\right) = 0,
\end{equation}
$\epsilon_{\tau\tau}$ and $|\epsilon_{e\tau}|$ satisfy a parabolic relation
\begin{equation}
\label{eq:ansatz_a}
\epsilon_{\tau\tau}=\frac{|\epsilon_{e\tau}|^2}{1+\epsilon_{ee}}
\end{equation}
and hence $\epsilon_{\tau\tau}$ can be eliminated. 
In the limit of Eq.\,(\ref{eq:ansatz_a}), the disappearance oscillation probability of the high-energy atmospheric neutrinos can be reduced to $\nu_{\mu} \leftrightarrow \nu_{\tau}'$ vacuum oscillation like two-flavor form ($\nu_{\tau}'$ is a mixture of $\nu_{e}$ and $\nu_{\mu}$ due to the presence of NSI) in spite of nonvanishing $\tau\tau$ component in the matter potential.
This means that the disappearance oscillation probability with NSI of the high-energy atmospheric neutrinos is proportional to $E^{-2}$
\begin{eqnarray}
1-P(\nu_{\mu} \rightarrow \nu_{\mu}) = \sin^22\theta_{\rm atm}\sin^2 \left(\frac{\Delta m^2_{\rm atm}L}{4E}\right) \propto \frac{1}{E^2}
\end{eqnarray}
as in the case of the standard two flavor neutrino oscillation framework.

Next let us introduce the matter angle $\beta$ \cite{Friedland:2004ah,Friedland:2005vy} which determines the mixing between the standard flavor basis $\nu_{e,\tau}$ defined by the W-boson exchange interaction and the modified flavor basis $\nu'_{e,\tau}$ due to the presence of NSI with components $\epsilon_{\alpha\beta}~(\alpha,\beta=e,\tau)$.
It is convenient to take the modified flavor basis in the discussion on the sensitivity of atmospheric neutrino experiments to NSI.
The matter angle $\beta$ is defined as
\begin{eqnarray}
\tan\beta 
\equiv \frac{|\epsilon_{e\tau}|}{1+\epsilon_{ee}}.
\label{tanb}
\end{eqnarray}
In the case of SK for 4438 days analysis, the constraint to $|\tan\beta|$ from the energy rate analysis is given by $|\tan\beta|\lesssim 0.8~(\mbox{\rm at}~2.5\sigma)$\,\cite{Fukasawa:2015jaa}.  If we rewrite the matter potential as
\begin{eqnarray}
{\cal A} = \sqrt{2} G_F N_e\left(
\begin{array}{ccc}
1+\epsilon_{ee}&0&\epsilon_{e\tau}\cr
0&0&0\cr
\epsilon_{e\tau}^\ast&0&|\epsilon_{e\tau}|^2/(1+\epsilon_{ee})
\end{array}\right)\,,
\nonumber
\end{eqnarray}
then the allowed region which was obtained in Ref.\,\cite{Fukasawa:2015jaa}
from the SK atmospheric neutrino data at $2.5\sigma$ is
\begin{eqnarray}
&{\ }&
\hspace*{-30mm}
-4 \lesssim \epsilon_{ee} \lesssim 4,
\nonumber\\
&{\ }&
\hspace*{-30mm}
|\epsilon_{e\tau}|\lesssim 3,
\nonumber\\
&{\ }&
\hspace*{-30mm}
|\epsilon_{\tau\tau}|=\frac{|\epsilon_{e\tau}|^2}{|1+\epsilon_{ee}|}
\lesssim 2\,.
\label{ett}
\end{eqnarray}
Notice that the bound (\ref{ett}) on $\epsilon_{\tau\tau}$
is much weaker than
what is obtained from the two flavor analysis assuming
only the $\nu_\mu\leftrightarrow\nu_\tau$
transition\,\cite{GonzalezGarcia:1998hj,Lipari:1999vh,Fornengo:1999zp,Fornengo:2001pm,GonzalezGarcia:2004wg,Mitsuka:2011ty}.  This is because in the two flavor
analysis $\epsilon_{ee}=\epsilon_{e\mu}=\epsilon_{e\tau}=0$ is assumed,
and the parabolic relation (\ref{eq:ansatz_a}) would
imply $\epsilon_{\tau\tau}\simeq 0$ in this case.\footnote{
Ref.\,\cite{Mitsuka:2011ty} also performed a
three flavor hybrid analysis with the $\epsilon_{\alpha\beta}~(\alpha,\beta=e,\tau)$
NSI components, and they obtained the bound
$|\epsilon_{\tau\tau}|\lesssim 0.15$ (at 90\%CL for $\epsilon_{ee}=1.5$)
which seems to be stronger than (\ref{ett}).
It is not clear whether the bounds on $|\epsilon_{\tau\tau}|$
in Ref.\,\cite{Mitsuka:2011ty} and in Ref.\,\cite{Fukasawa:2015jaa} are consistent with each other,
since the analysis in Ref.\,\cite{Mitsuka:2011ty}
(full information of the zenith angle and energy spectral bins is taken
into account while the phase of $\epsilon_{e\tau}$ is
not taken into consideration and they studied only the region 
$-1.5 \le \epsilon_{ee} \le 1.5$) is
different from that in Ref.\,\cite{Fukasawa:2015jaa}
(the phase of $\epsilon_{e\tau}$ is taken into account and
the wider region $-4 \le \epsilon_{ee} \le 4$ was studied while
the energy spectral information is not taken into consideration).}

It is instructive to discuss the relation between the standard parametrization
$\epsilon_{\alpha\beta}$ and the set of the parametrizations
($\epsilon^f_{D}$, $\epsilon^f_{N}$) in the simplest case.
In the simplest case, we assume
the parabolic relation (\ref{eq:ansatz_a})
and set $\theta_{13}=0$, $\theta_{23}=\pi/4$,
which is a good approximation to some extent.
Then, introducing a new angle
\begin{eqnarray}
&{\ }&
\hspace*{-80mm}
\tan\beta'\equiv\frac{\tan\beta}{\sqrt{2}}\,,
\label{tanbprime}
\end{eqnarray}
we can derive the following relation
(See Appendix\,\ref{appendixa} for the derivation
and the expression for a more general case.):
\begin{eqnarray}
&{\ }&
\hspace*{-70mm}
\frac{\left|3\epsilon_{N}\right|}
{1/2-3\epsilon_{D}}
=\tan2\beta'\,.
\label{gradbetaprime}
\end{eqnarray}
The region
$|\epsilon_{e\tau}|/|1+\epsilon_{ee}|<\tan\beta$\,,
which is the area surrounded by the $\epsilon_{e\tau}=0$ axis
and the straight line $|\epsilon_{e\tau}|= \tan\beta\,|1+\epsilon_{ee}|$
with the gradient $\tan\beta$ and the $x$-intercept $\epsilon_{ee}=-1$,
is the allowed region in the ($\epsilon_{ee}$, $|\epsilon_{e\tau}|$)
plane by the atmospheric neutrino data under the assumption of the
parabolic relation (\ref{eq:ansatz_a}).
The corresponding region in the ($\epsilon_{D}$, $\epsilon_{N}$)
plane is approximately given by the one surrounded by the $\epsilon_{N}=0$ axis
and the straight line
$|\epsilon_{N}|= \tan2\beta'\,|1/6-\epsilon_{D}|$
with the gradient $\tan2\beta'$ and the $x$-intercept $\epsilon_{D}=1/6$.

\subsubsection{The case with $\epsilon_{\alpha\mu}\ne 0~(\alpha=e,\mu,\tau)$}
From here we take into consideration
all the components of $\epsilon_{\alpha\beta}$
including the $\mu$ components, and lift the 
parabolic relation (\ref{eq:ansatz_a}).
Even in this case, because of the strong constraints (\ref{epsilon-m})
on the $\epsilon_{\alpha\mu}$ components, the three
eigenvalues of the matter potential matrix ${\cal A}$
are approximately 0 and
$1+\epsilon_{ee}+\epsilon_{\tau\tau}\pm\sqrt{(1+\epsilon_{ee}-\epsilon_{\tau\tau})^2+4|\epsilon_{e\tau}|^2}$.
So most of the discussions in the previous subsubsection
are approximately valid.  In particular,
the constraint from the high energy data of the
atmospheric neutrinos can be approximately given by
Eq.\,(\ref{eq:ansatz_b}).
We note that another derivation of
the relation (\ref{eq:ansatz_a}) was given in Ref.\,\cite{Oki:2010uc}.
The high-energy behavior of the disappearance oscillation probability in the presence of NSI without switching off any $\epsilon_{\alpha\beta}$ can be written as
\begin{eqnarray}
1-P(\nu_{\mu} \rightarrow \nu_{\mu}) \simeq c_0+ c_1\frac{\sqrt{2}G_FN_e}{E}+ {\cal O} \left(\frac{1}{E^2} \right).
\end{eqnarray}
This expression requires $|c_0| \ll 1$ and $|c_1| \ll 1$ so that the presence of NSI is consistent with the high-energy atmospheric neutrino experiments data.
The constraints on $c_0$ and $c_1$ imply $\epsilon_{e\mu}\simeq\epsilon_{\mu\mu}\simeq\epsilon_{\tau\mu}\simeq0$ and $\epsilon_{\tau\tau}\simeq|\epsilon_{e\tau}|^2/(1+\epsilon_{ee})$.

\section{Analysis\label{analysis}}
In this section we discuss the sensitivity of the Hyper-Kamiokande (HK) atmospheric neutrino experiment whose data is assumed to be taken for 4438 days to $\epsilon_{D}$ and $|\epsilon_{N}|$ with the codes that were used in Ref.\,\cite{Foot:1998iw,Yasuda:1998mh,Yasuda:2000de,Fukasawa:2015jaa}.
We assume that the HK fiducial volumes are 0.56 Mton\,\footnote{
Recently there is a new design with the reduced fiducial volume (two tanks
with the fiducial volume 0.19 Mton each).
However, decreasing of the number of events due to a reduced fiducial volume can be compensated by the improvement of the detection efficiencies.  Since the details of the new design are not known,  we will
analyze the atmospheric neutrino
measurements at Hyperkamiokande
with the parameters in the
old design
throughout this paper.
}, and that the HK detector has the same detection efficiencies as those of Super-Kamiokande (SK) and that HK atmospheric neutrino data comprise the sub-GeV, multi-GeV and upward going $\mu$ events as in the case of SK.
As HK is the future experiment, the number of events calculated with the standard three flavor oscillation scenario are used as the experimental data for fitting.
The reference values of oscillation parameters used in the calculation of the experimental data are the following:
\begin{eqnarray}
\Delta \bar{m}^2_{31}=2.5\times 10^{-3}\mbox{\rm eV}^2,
\sin^2\bar{\theta}_{23}=0.5,
\bar{\delta}_{\rm CP}=0,
\nonumber\\
\sin^22\bar{\theta}_{12}=0.86,
\sin^22\bar{\theta}_{13}=0.1,
\Delta \bar{m}^2_{21}=7.6\times 10^{-5}\mbox{\rm eV}^2\,,
\label{ref-value}
\end{eqnarray}
where the parameters with a bar denote
those for the reference value of ``the experimental data''.
The information on the zenith angle bins for the sub-GeV, multi-GeV and upward going $\mu$ events are given in Ref.\,\cite{Abe:2014gda} while that on the energy bins is not.
We analyze with the ten zenith angle bins as in Ref.\,\cite{Abe:2014gda}.
As the experimental data is calculated by our codes, we can use any information on the energy spectrum of the number of events and analyze with any number of the energy bins.

The analysis was performed using $\chi^2$-method and $\chi^2$  is defined as
\begin{eqnarray}
\chi^2=
\min_{\theta_{23},|\Delta m^2_{32}|,\delta,\epsilon_{\alpha\beta}}
\left(
\chi_{\rm sub-GeV}^2+\chi_{\rm multi-GeV}^2
+\chi_{\rm upward}^2
+\chi_{\rm prior}^2\right),
\label{eqn:chi}
\end{eqnarray}
where
\begin{eqnarray}
\hspace*{-20mm}
&{\ }&
\displaystyle\chi_{\rm sub-GeV}^2
\nonumber\\
&=&
\min_{\alpha_s,\beta's,\gamma's}\left[
\frac{\beta_{s1}^2}{\sigma_{\beta s1}^2}
+\frac{\beta_{s2}^2}{\sigma_{\beta s2}^2}
+\frac{\gamma_{L1}^2}{\sigma_{\gamma L1}^2}
+\frac{\gamma_{L2}^2}{\sigma_{\gamma L2}^2}
+\frac{\gamma_{H1}^2}{\sigma_{\gamma H1}^2}
+\frac{\gamma_{H2}^2}{\sigma_{\gamma H2}^2}
\right.
\nonumber\\
&{\ }&\quad
+\sum_{A=L,H}\sum_{j=1}^{10}\left\{
\frac{1}{n_{Aj}^s(e)}
\left[ \alpha_s \left(1-{\beta_{s1} \over 2}+{\beta_{s2}
\over 2} +  {\gamma_{A1}^j \over 2}  \right)N_{Aj}^s(\nu_e\to\nu_e)
\right.\right.
\nonumber\\
&{\ }&\quad
+ \alpha_s \left(1+{\beta_{s1} \over 2}+{\beta_{s2}
\over 2}+{\gamma_{A1}^j \over 2}\right)N_{Aj}^s(\nu_\mu\to\nu_e)
\nonumber\\
&{\ }&\quad+ \alpha_s \left(1-{\beta_{s1} \over 2}-{\beta_{s2}
\over 2}+{\gamma_{A1}^j \over 2}\right)N_{Aj}^s(\bar{\nu}_e\to\bar{\nu}_e)
\nonumber\\
&{\ }&\quad\left.
+ \alpha_s \left(1+{\beta_{s1} \over 2}-{\beta_{s2}
\over 2}+{\gamma_{A1}^j \over 2}\right)N_{Aj}^s(\bar{\nu}_\mu\to\bar{\nu}_e)
-n_{Aj}^s(e)\right]^2
\nonumber\\
&{\ }&\quad
+\frac{1}{n_{Aj}^s(\mu)} \left[ 
\alpha_s \left(1-{\beta_{s1} \over 2}+{\beta_{s2} 
\over 2}+{\gamma_{A2}^j \over 2}\right)N_{Aj}^s(\nu_e\to\nu_\mu)
\right.\nonumber\\
&{\ }&\quad
+\alpha_s \left(1+{\beta_{s1} \over 2}+{\beta_{s2} 
\over 2}+{\gamma_{A2}^j \over 2}\right)N_{Aj}^s(\nu_\mu\to\nu_\mu)
\nonumber\\
&{\ }&\quad+\alpha_s \left(1-{\beta_{s1} \over 2}-{\beta_{s2} 
\over 2}+{\gamma_{A2}^j \over 2}\right)N_{Aj}^s(\bar{\nu}_e\to\bar{\nu}_\mu)
\nonumber\\
&{\ }&\quad\left.\left.\left.+\alpha_s \left(1+{\beta_{s1} \over 2}-{\beta_{s2} 
\over 2}+{\gamma_{A2}^j \over 2}\right)N_{Aj}^s(\bar{\nu}_\mu\to\bar{\nu}_\mu)
-n_{Aj}^s(\mu)\right]^2
\right\}\right],
\label{eqn:chi-sub2}
\end{eqnarray}
\begin{eqnarray}
&{\ }&\displaystyle\chi_{\rm multi-GeV}^2\nonumber\\
&=&
\min_{\alpha_m,\beta's,\gamma's}\left[
\frac{\beta_{m1}^2}{\sigma_{\beta m1}^2}
+\frac{\beta_{m2}^2}{\sigma_{\beta m2}^2}
+\frac{\gamma_{1}^2}{\sigma_{\gamma 1}^2}
+\frac{\gamma_{2}^2}{\sigma_{\gamma 2}^2}\right.
\nonumber\\
&{\ }&+\sum_{A=L,H}\sum_{j=1}^{10}\left\{
\frac{1}{n_{Aj}^m(e)}
\left[ \alpha_m \left(1-{\beta_{m1} \over 2}+{\beta_{m2}
\over 2}+{\gamma_{1}^j \over 2} \right)N_{Aj}^m(\nu_e\to\nu_e)
\right.\right.
\nonumber\\
&{\ }&+ \alpha_m \left(1+{\beta_{m1} \over 2}+{\beta_{m2}
\over 2}+{\gamma_{1}^j \over 2}\right)N_{Aj}^m(\nu_\mu\to\nu_e)
\nonumber\\
&{\ }&+ \alpha_m \left(1-{\beta_{m1} \over 2}-{\beta_{m2}
\over 2}+{\gamma_{1}^j \over 2}\right)N_{Aj}^m(\bar{\nu}_e\to\bar{\nu}_e)
\nonumber\\
&{\ }&\left.+ \alpha_m \left(1+{\beta_{m1} \over 2}-{\beta_{m2}
\over 2}+{\gamma_{1}^j \over 2}\right)N_{Aj}^m(\bar{\nu}_\mu\to\bar{\nu}_e)
-n_{Aj}^m(e)\right]^2
\nonumber\\
&{\ }&+\frac{1}{n_{Aj}^m(\mu)}
\left[ 
\alpha_m \left(1-{\beta_{m1} \over 2}+{\beta_{m2} 
\over 2}+{\gamma_{2}^j \over 2}\right)N_{Aj}^m(\nu_e\to\nu_\mu)\right.
\nonumber\\
&{\ }&+\alpha_m \left(1+{\beta_{m1} \over 2}+{\beta_{m2} 
\over 2}+{\gamma_{2}^j \over 2}\right)N_{Aj}^m(\nu_\mu\to\nu_\mu)
\nonumber\\
&{\ }&+\alpha_m \left(1-{\beta_{m1} \over 2}-{\beta_{m2} 
\over 2}+{\gamma_{2}^j \over 2}\right)N_{Aj}^m(\bar{\nu}_e\to\bar{\nu}_\mu)
\nonumber\\
&{\ }&\left.\left.\left.+\alpha_m \left(1+{\beta_{m1} \over 2}-{\beta_{m2} 
\over 2}+{\gamma_{2}^j \over 2}\right)N_{Aj}^m(\bar{\nu}_\mu\to\bar{\nu}_\mu)
-n_{Aj}^m(\mu)\right]^2
\right\}\right],
\label{eqn:chi-multi2}
\end{eqnarray}
\begin{eqnarray}
\displaystyle\chi_{\rm upward}^2
&=&
\min_{\alpha_u}\left\{
\frac{\alpha_u^2}{\sigma_{\alpha}^2}
+\sum_{j=1}^{10}
\frac{1}{n_j^u(\mu)}
\left[ 
(1+\alpha_u) N_j^u(\nu_e\to\nu_\mu)
+(1+\alpha_u) N_j^u(\nu_\mu\to\nu_\mu)\right.\right.
\nonumber\\
&{\ }&\qquad\left.\left.+(1+\alpha_u) N_j^u(\bar{\nu}_e\to\bar{\nu}_\mu)
+(1+\alpha_u) N_j^u(\bar{\nu}_\mu\to\bar{\nu}_\mu)
-n_j^u(\mu)\right]^2\right\},
\end{eqnarray}
\begin{eqnarray}
\displaystyle\chi_{\rm prior}^2
&=& \Delta\chi^2_{\rm prior}  \frac{|\epsilon^f_{e\mu}|^2}{|\delta \epsilon^f_{e\mu}|^2}
+\Delta\chi^2_{\rm prior} \frac{|\epsilon^f_{\mu\tau}|^2}{|\delta \epsilon^f_{\mu\tau}|^2} .
\end{eqnarray}
Where $\Delta\chi^2_{\rm prior}=2.7$ in $\chi_{\rm prior}^2$ stands for $\chi^2$ for $90\%$CL with 1 d.o.f. and $|\delta \epsilon^f_{e\mu}|=|\delta \epsilon^f_{\mu\tau}|=0.05$ stand for constraint on corresponding NSI at $90\%$CL, respectively.
The summation on $j$  and  $A=L,H$ run over the ten zenith angle bins and the two energy bins, respectively.
The indices $L$ and $H$ stand for the lower ($E<E_{\mbox{\rm\scriptsize th}}$) and higher ($E>E_{\mbox{\rm\scriptsize th}}$) energy bins, respectively.
For all the zenith angle bins, the threshold energy for the sub-GeV events is 0.5GeV and that for the multi-GeV events is 3.2GeV.
The threshold energy $E_{\mbox{\rm\scriptsize th}}$ is chosen so that the numbers of events for the lower and higher energy bins are approximately equal.
The experimental data $n_{Aj}^a(\alpha)$ ($a=s,m; \alpha=e,\mu$) stands for the sum of the number of neutrinos and antineutrinos events for the sub-GeV and multi-GeV events, and the experimental data $n_j^u(\mu)$ stands for that for the upward going $\mu$ events.
$N_{Aj}^a(\nu_\alpha\to\nu_\beta)$($N_{Aj}^a(\bar{\nu}_\alpha\to\bar{\nu}_\beta)$) stands for the prediction with our codes for the number of $\ell_{\beta}$-like events ($\ell_{\beta}=e,\mu$) of the sub-GeV and multi-GeV events and $N_{j}^u(\nu_\alpha\to\nu_\beta)$($N_{j}^a(\bar{\nu}_\alpha\to\bar{\nu}_\beta)$) stands for that of the upward going $\mu$ events. 
$\alpha_a~(a=s, m, u)$ stands for the uncertainty in the overall flux normalization for the sub-GeV, multi-GeV, and upward going $\mu$ events, $\beta_{a1}$ ($\beta_{a2}$) stands for the uncertainty in the relative normalization between $\nu_e$ - $\nu_\mu$ flux ($\nu$ - $\bar{\nu}$ flux) for the sub-GeV ($a=s$) and multi-GeV ($a=m$) events, respectively, and $\gamma's$ stand for the flavor and energy dependent relative normalization between the upward and downward bins for the sub-GeV and multi-GeV events:
\begin{eqnarray}
\gamma_{A1,2}^j&=&
\left\{ \begin{array}{ll}
\gamma_{A1,2} & (j \le j_{\mbox{\rm\scriptsize th}}; A=L, H) \nonumber\\
-\gamma_{A1,2} & (j > j_{\mbox{\rm\scriptsize th}}; A=L, H) \nonumber\\
\end{array} \right.\nonumber\\
\gamma_{1,2}^j&=&
\left\{ \begin{array}{ll}
\gamma_{1,2} & (j \le j_{\mbox{\rm\scriptsize th}}) \nonumber\\
-\gamma_{1,2} & (j > j_{\mbox{\rm\scriptsize th}}). \nonumber\\
\end{array} \right.
\end{eqnarray}
Here $j_{\mbox{\rm\scriptsize th}}=3$ is the index which separates the upward and downward bins and determined in the investigation of the significance of the wrong mass hierarchy.
We have checked that the choice of the upward-downward separation index $j_{\mbox{\rm\scriptsize th}}$ do not affect the sensitivity to NSI significantly. We have set the systematic errors to the same values as in Ref.\,\cite{Ashie:2005ik}
\begin{eqnarray}
&{\ }&\sigma_{\beta s1}=\sigma_{\beta m1} = 0.03, \nonumber\\
&{\ }&\sigma_{\beta s2} =\sigma_{\beta m2} = 0.05, \nonumber\\
&{\ }&\sigma_{\alpha} = 0.2, \nonumber\\
&{\ }&\sigma_{\gamma L1}=0.005, \nonumber\\
&{\ }&\sigma_{\gamma L2}=0.008, \nonumber\\
&{\ }&\sigma_{\gamma H1}=0.021, \nonumber\\
&{\ }&\sigma_{\gamma H2}=0.018, \nonumber\\
&{\ }&\sigma_{\gamma 1}=0.015, \nonumber\\
&{\ }&\sigma_{\gamma 2}=0.008,
\label{sys3}
\end{eqnarray}
and omitted other systematic errors for simplicity. In particular, we confirmed that taking a uncertainty in the $E_{\nu}$ spectral index which is omitted in our analysis into consideration gives negligible contributions to $\chi^2$.

Before moving on to the discussions on the sensitivity of HK to NSI, we investigate the significance of the wrong mass hierarchy to check the validity of our codes.
The significance of the wrong mass hierarchy is calculated with different numbers of the energy bins. 
We found that the significance calculated by our codes with two energy bins in the contained events and one energy bin in the upward going $\mu$ events is similar to the one in Ref.\,\cite{Abe:2011ts}.
The more we increase the energy bins, the larger significance of the wrong mass hierarchy becomes.
In the case of the analysis of the sensitivity to NSI, the allowed regions with more than two energy bins are smaller than those with two energy bins.
In this paper, therefore, we adopt two energy bins in the contained events and one energy bin in the upward going $\mu$ events so that the results are conservative.

The  sensitivity of the atmospheric neutrino experiment to NSI which is parametrized as ($\epsilon_{D}$, $\epsilon_{N}$) is studied as follows.
\begin{enumerate}
  \item Set a grid on the ($\epsilon_{D}$, $|\epsilon_{N}|$) plane.
  \item Calculate a parameter set $(|\epsilon_{e\tau}|,\epsilon_{ee}-\epsilon_{\mu \mu},\epsilon_{\tau\tau}-\epsilon_{\mu \mu})$ via Eq.\,(\ref{eq_nsi1}) for the given point ($\epsilon_{D}$, $|\epsilon_{N}|$) on the grid varying  $\Delta m^2_{31}$, $\theta_{23}$, $\delta_{\rm CP}$, $|\epsilon_{e\mu}|$, $|\epsilon_{\mu\tau}|$, $\psi$ and $\phi_{\alpha\beta}$.
  \item Dismiss the parameter set if it does not satisfy any one of the following criteria:    
  \begin{eqnarray}
|\epsilon_{e\tau}| \le 1.5 
\label{criterion1}\\
|\epsilon_{ee}-\epsilon_{\mu\mu}| \le 2.0
\label{criterion2}\\
\min_{\pm}
\left(\left|
1+(\epsilon_{ee}-\epsilon_{\mu\mu})+(\epsilon_{\tau\tau}-\epsilon_{\mu\mu})
\pm\sqrt{(1+\epsilon_{ee}-\epsilon_{\tau\tau})^2+4|\epsilon_{e\tau}|^2}\right|\right) \le 0.4
\label{criterion3}
\end{eqnarray}
  \item Calculate $\chi^2$ for each parameter set which passed the criteria mentioned above and then obtain the minimum value of $\chi^2$ for the given ($\epsilon_{D}$, $|\epsilon_{N}|$).
\end{enumerate}
As mentioned in subsection \ref{atm_nu}, the atmospheric neutrino experiments constrain the relation between $\epsilon_{ee}$, $|\epsilon_{e\tau}|$ and $\epsilon_{\tau\tau}$.
Eq.\,(\ref{criterion3}) is still valid when we replace $\epsilon_{e e}$  with $(\epsilon_{e e}-\epsilon_{\mu \mu})$ and $\epsilon_{\tau \tau}$  with $(\epsilon_{\tau\tau}-\epsilon_{\mu \mu})$.
This replacement can be understood as a redefinition of the standard NSI parameterization because one can always subtract the modified MSW potential (\ref{matter-np}) by a matrix proportional to identity, say $\epsilon_{\mu\mu} \mathbf{1}_{3\times3}$, as far as the oscillation probability is concerned.
Therefore if the parameter set $(|\epsilon_{e\tau}|,\epsilon_{ee}-\epsilon_{\mu \mu},\epsilon_{\tau\tau}-\epsilon_{\mu \mu})$ which is determined by the independent parameters dose not satisfy Eq.\,(\ref{criterion3}), we can exclude it without fitting to the experimental data.
In addition to this criterion, we can also exclude the parameter set which dose not satisfy Eq.\,(\ref{criterion1}) or Eq.\,(\ref{criterion2}).
The criterions Eqs.\,(\ref{criterion1}) and (\ref{criterion2}) are justified by the results in the previous work\,\cite{Fukasawa:2015jaa}.

In our analysis of the sensitivity to NSI, we assume that the mass hierarchy is known because there may be some hints on the mass hierarchy determination by the time HK accumulate the data for 4438 days.
Variations in $\Delta m^2_{21}$, $\theta_{12}$ and $\theta_{13}$ give little effect on the sensitivity of HK to NSI, and hence we fix them as the same values of the experimental data in fitting.
Taking into account the constraints on NSI given by the previous researches, we vary NSI parameters as follows:
\begin{eqnarray}
0\leq &|\epsilon_{e\mu}^d|& \leq 0.05 \nonumber \\
0\leq &|\epsilon_{\mu\tau}^d|& \leq 0.05 \nonumber \\
0\leq &\phi_{\alpha\beta}& < 2\pi \nonumber \\
0\leq &\psi& < 2\pi.
\end{eqnarray}

\begin{figure}[H]
\includegraphics[scale=0.4,angle=-90]{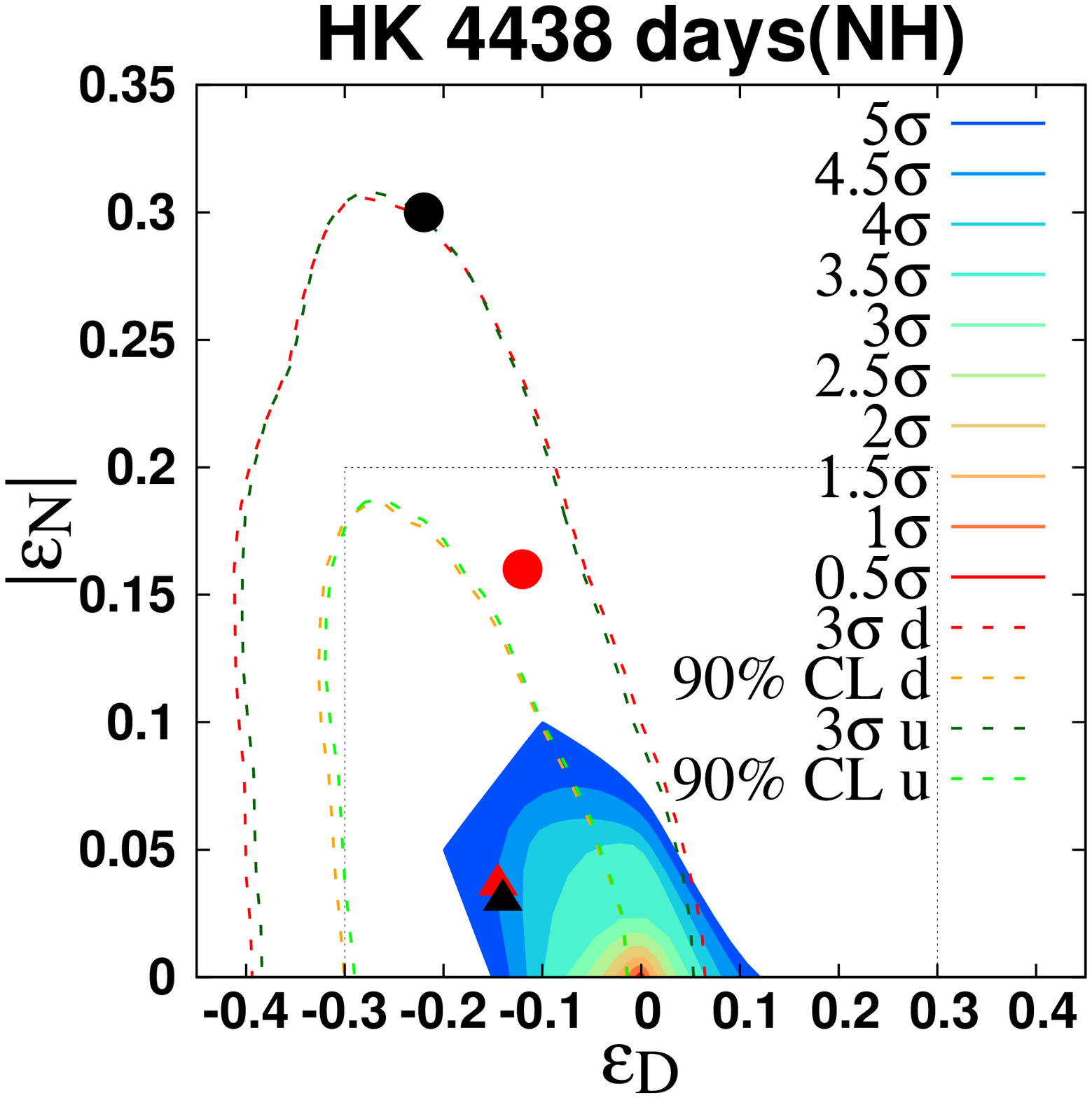}
\includegraphics[scale=0.4,angle=-90]{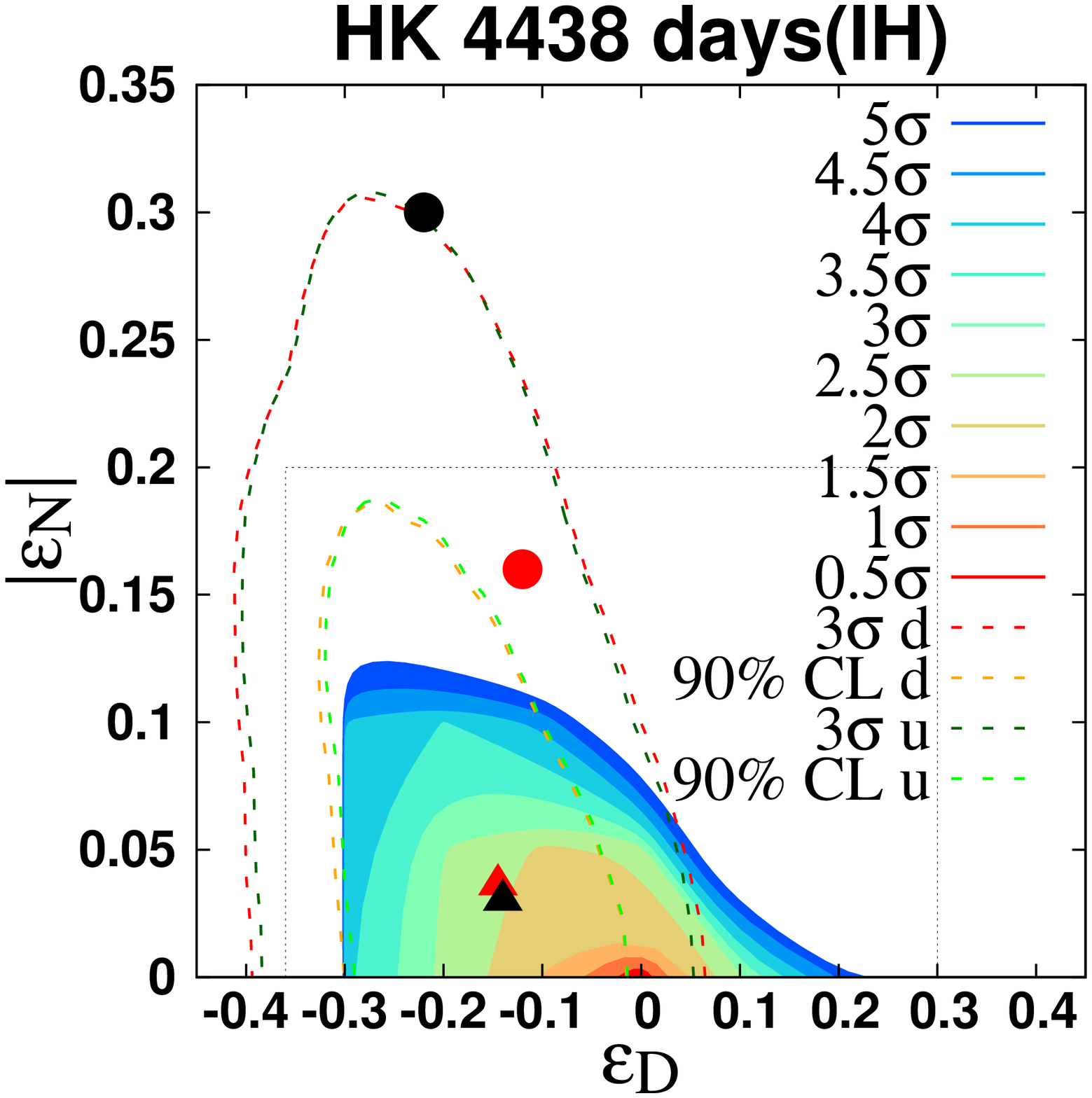}
\caption{The allowed region in the ($\epsilon_{D}$, $|\epsilon_{N}|$) plane from the HK atmospheric neutrino data for the normal hierarchy (left panel) and for the inverted hierarchy (right panel).
We calculated $\chi^2$ for ($\epsilon_{D}$, $|\epsilon_{N}|$) inside the area surrounded by dotted lines and at the best fit points. The red ($f=d$) and black ($f=u$) circles indicate the best fit point from the solar neutrino and KamLAND data \cite{Gonzalez-Garcia:2013usa} for NSI with $(\epsilon_{D}^d,\epsilon_{N}^d)=(-0.12,-0.16)$ (red) and that for NSI with $(\epsilon_{D}^u,\epsilon_{N}^u)=(-0.22,-0.30)$ (black), respectively.
In the case of the normal hierarchy, $\chi^2$ for the red and black circles are 128.49 (11$\sigma$) and 1670.4 (38$\sigma$), respectively, and in the case of the inverted hierarchy, $\chi^2$ for the red and black circles are 72.531 (8.2$\sigma$) and 1265.4 (35$\sigma$), respectively.
The red and black triangles indicate the best fit value from the global neutrino oscillation experiments analysis \cite{Gonzalez-Garcia:2013usa} for NSI with $(\epsilon_{D}^d,\epsilon_{N}^d)=(-0.145,-0.036)$ (red) and that for NSI with $(\epsilon_{D}^u,\epsilon_{N}^u)=(-0.140,-0.030)$ (black), respectively.
In the case of the normal hierarchy, $\chi^2$ for the red and black triangles are 28.967 (5.0$\sigma$) and 28.2934 (5.0$\sigma$), respectively, and in the case of the inverted hierarchy, $\chi^2$ for the red and black triangles are 4.1077 (1.5$\sigma$) and 3.7412 (1.4$\sigma$), respectively.
The dashed lines are the boundaries of the allowed regions from the global neutrino oscillation experiments analysis.
For reference, we plotted for both the cases with $f=u$ and $f=d$.} 
\label{fig:fig1}
\end{figure}

The results are shown in Fig.\,\ref{fig:fig1}.
The best fit values $(\epsilon_{D}^d,\epsilon_{N}^d)=(-0.12,-0.16)$ for NSI with $f=d$ from the solar neutrino and KamLAND data given by Ref.\,\cite{Gonzalez-Garcia:2013usa} is excluded at $11\sigma$ ($8.2\sigma$) for the normal (inverted) hierarchy.
In the case of NSI with $f=u$, the best fit value $(\epsilon_{D}^u,\epsilon_{N}^u)=(-0.22,-0.30)$ is far from the standard scenario $(\epsilon_{D},\epsilon_{N})=(0.0,0.0)$ compared with the case of $f=u$ and also excluded at $38\sigma$ ($35\sigma$) for the normal (inverted) hierarchy.
On the other hand, the best fit value from the global analysis of the neutrino oscillation data \cite{Gonzalez-Garcia:2013usa} 
$(\epsilon_{D}^d,\epsilon_{N}^d)=(-0.145,-0.036)$ for NSI with $f=d$
is excluded at $5.0\sigma$ ($3.7\sigma$)  for the normal (inverted) hierarchy.
In the case of NSI with $f=u$, the best fit value 
$(\epsilon_{D}^u,\epsilon_{N}^u)=(-0.140,-0.030)$
is excluded at $5.0\sigma$ ($1.4\sigma$) 
for the normal (inverted) hierarchy.
Notice that the fermion subscript $f$ on  $\epsilon_{D}^f$ and $\epsilon_{N}^f$ is important in the case of the solar neutrinos analysis because the number densities of up and down quarks are different in the Sun.
On the other hand, as mentioned above, in the case of atmospheric neutrinos the fermion subscript is not important because the number densities of up and down quarks are approximately the same in the Earth.

\begin{figure}[H]
\includegraphics[scale=0.4,angle=-90]{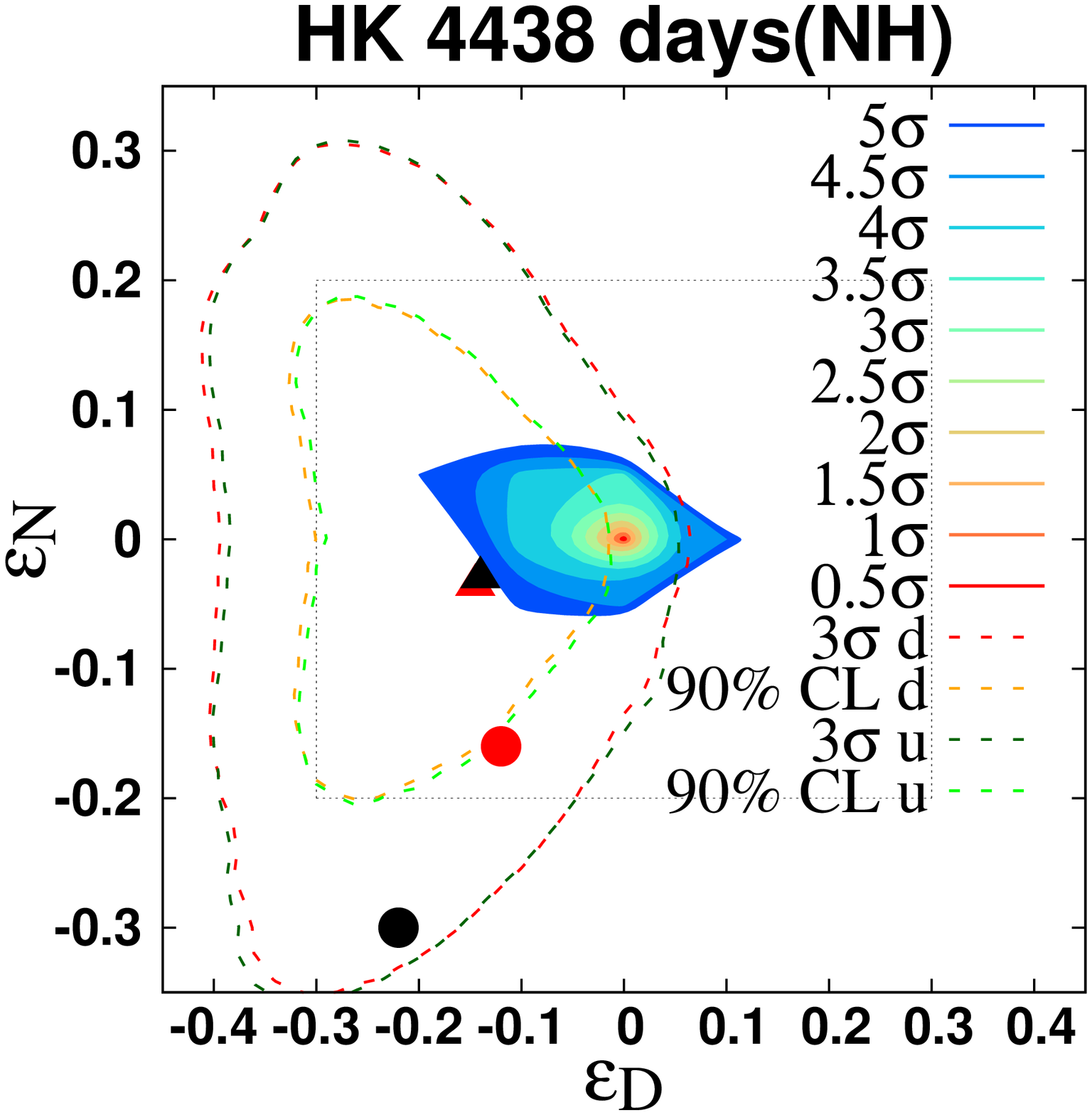}
\includegraphics[scale=0.4,angle=-90]{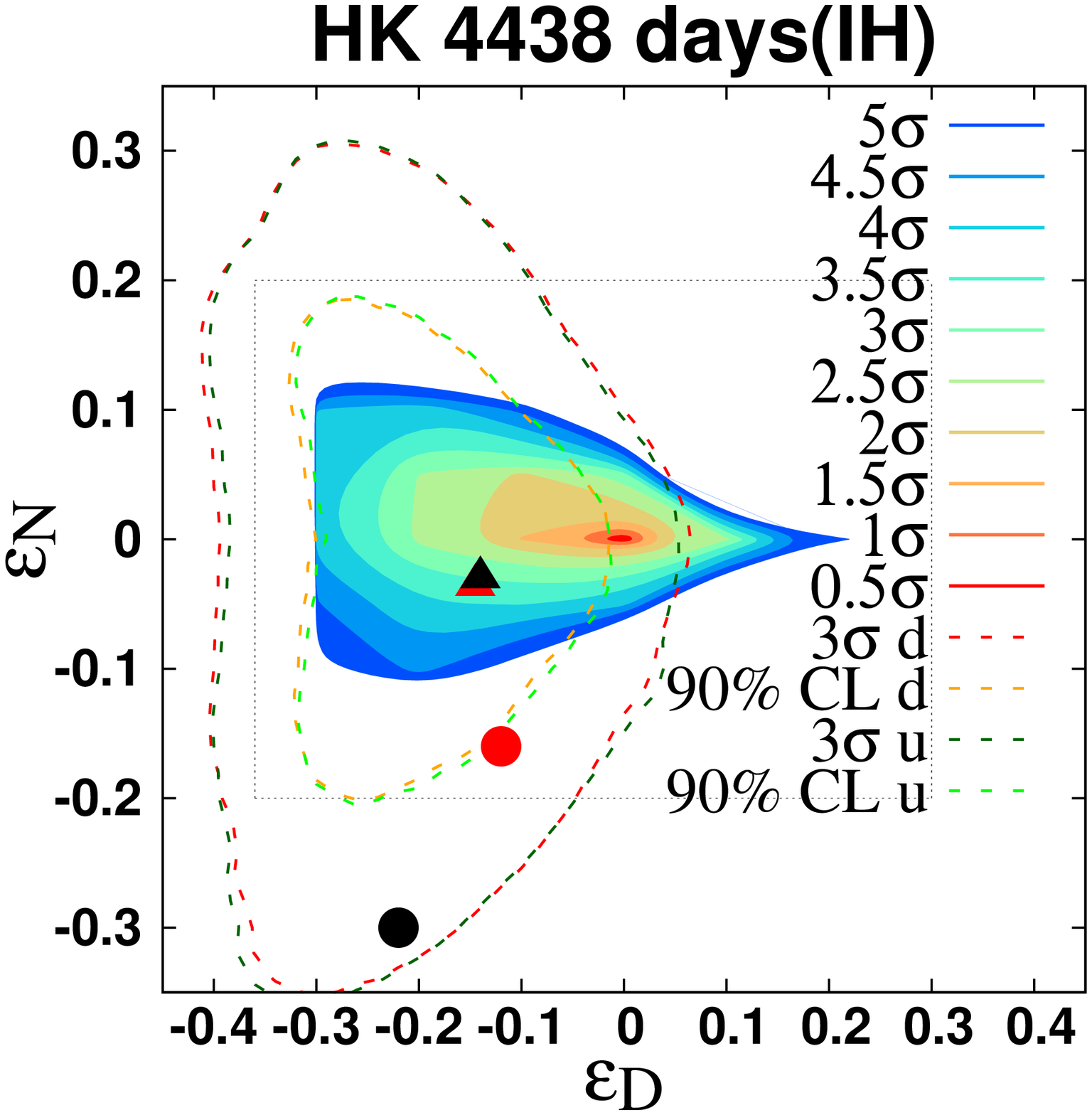}
\caption{The allowed regions assuming real $\epsilon_N$.} 
\label{fig:fig1_1}
\end{figure}

To compare our results with the one given in Ref.\,\cite{Gonzalez-Garcia:2013usa}, we show the allowed regions assuming real $\epsilon_N$ in Fig.\,\ref{fig:fig1_1}.
This is given by setting $\psi=0,\pi$ in Eq.\,(\ref{epsilonn}), where $\delta_{\rm CP}$ and $\phi_{\alpha\beta}$ do not need to be $0$ or $\pi$.
As can be seen from Fig.\,\ref{fig:fig1_1}, the allowed regions for positive $\epsilon_N$ and for negative one are almost symmetric.
We found that the allowed regions in Fig.\,\ref{fig:fig1} are not so different from that in the upper plane of Fig.\,\ref{fig:fig1_1}.
Therefore the sensitivity of the HK atmospheric neutrino experiment to $\psi$ is poor.

\begin{figure}[H]
\includegraphics[scale=0.3,angle=-90]{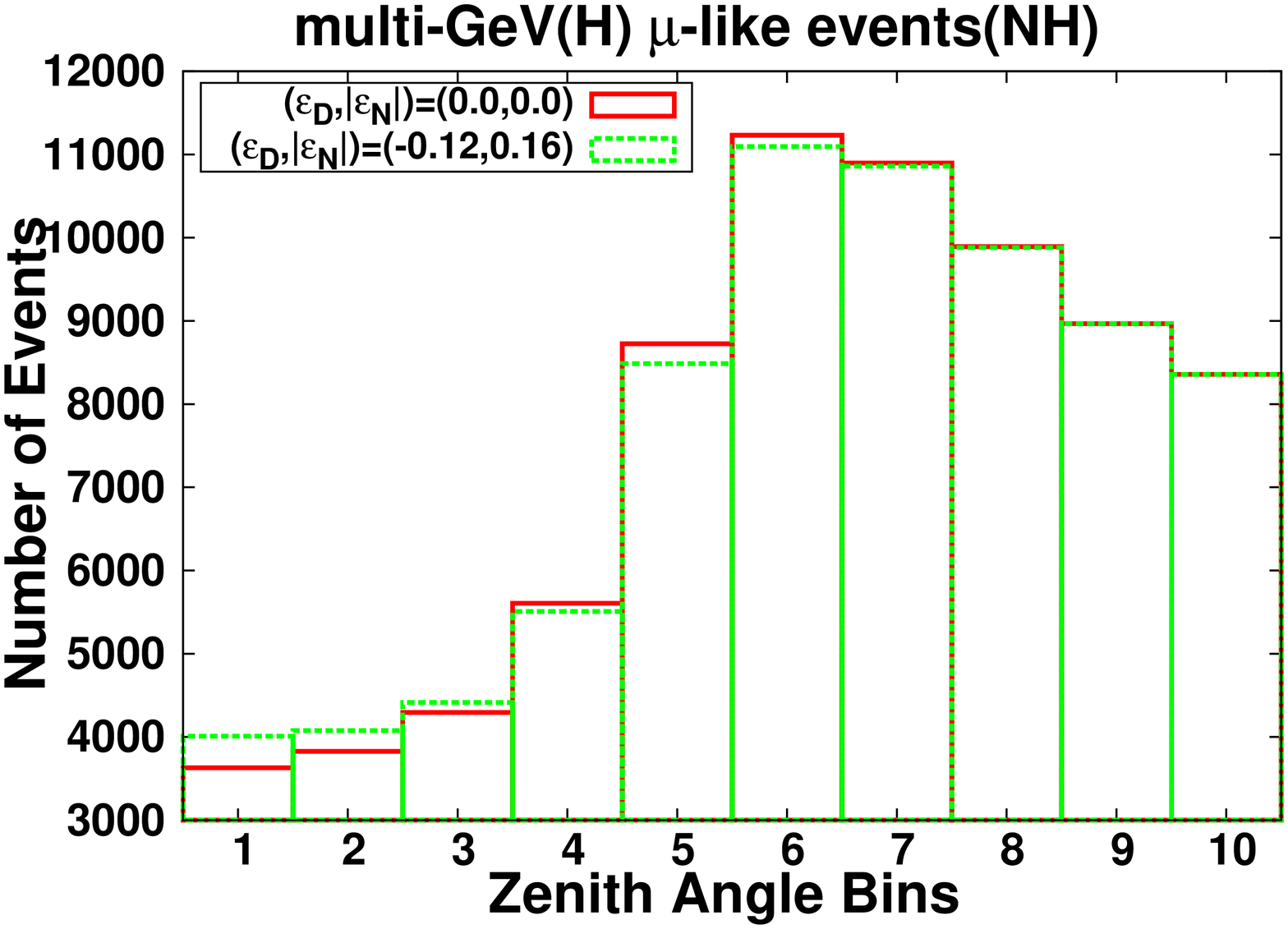}
\includegraphics[scale=0.3,angle=-90]{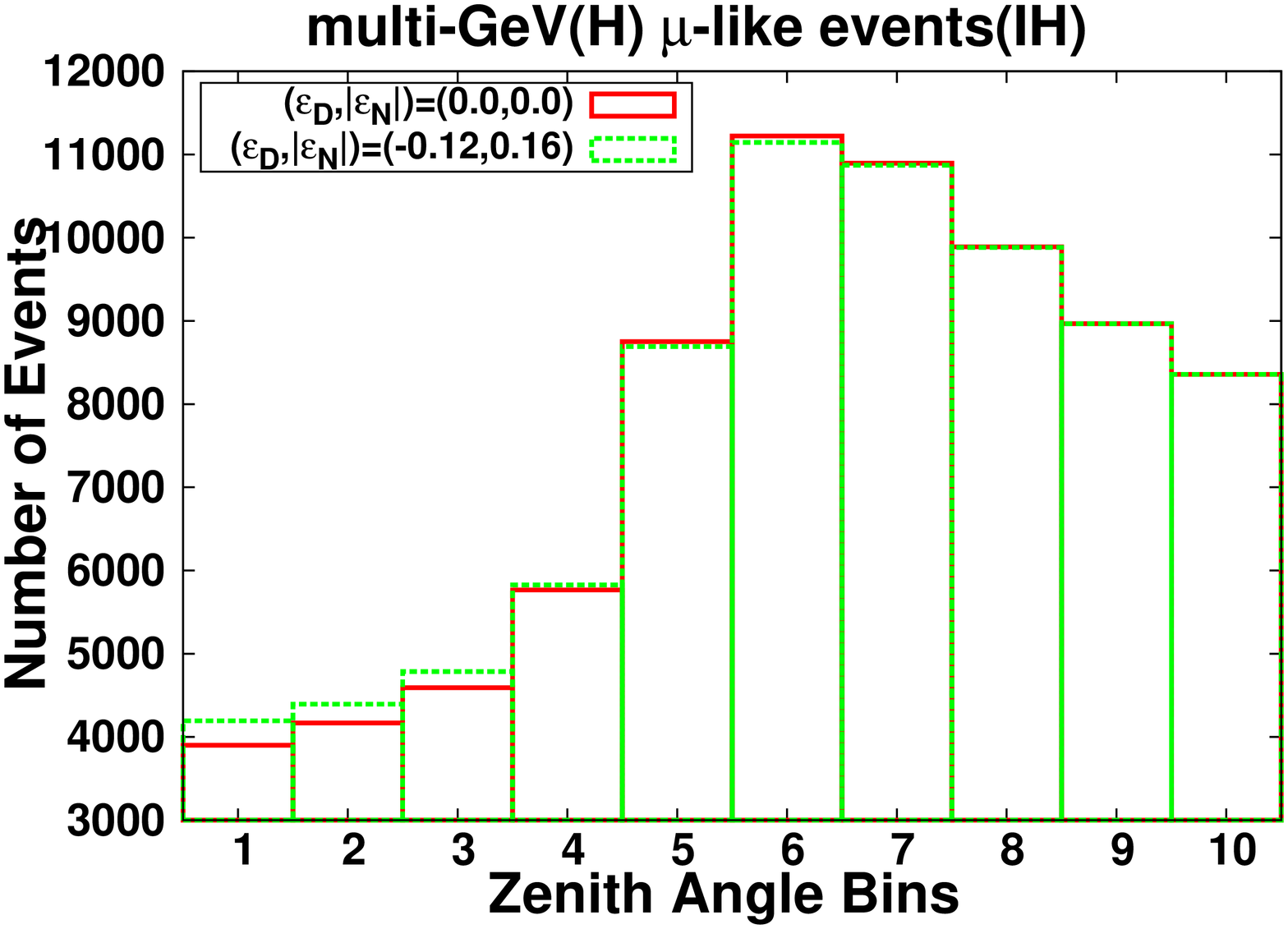}
\includegraphics[scale=0.3,angle=-90]{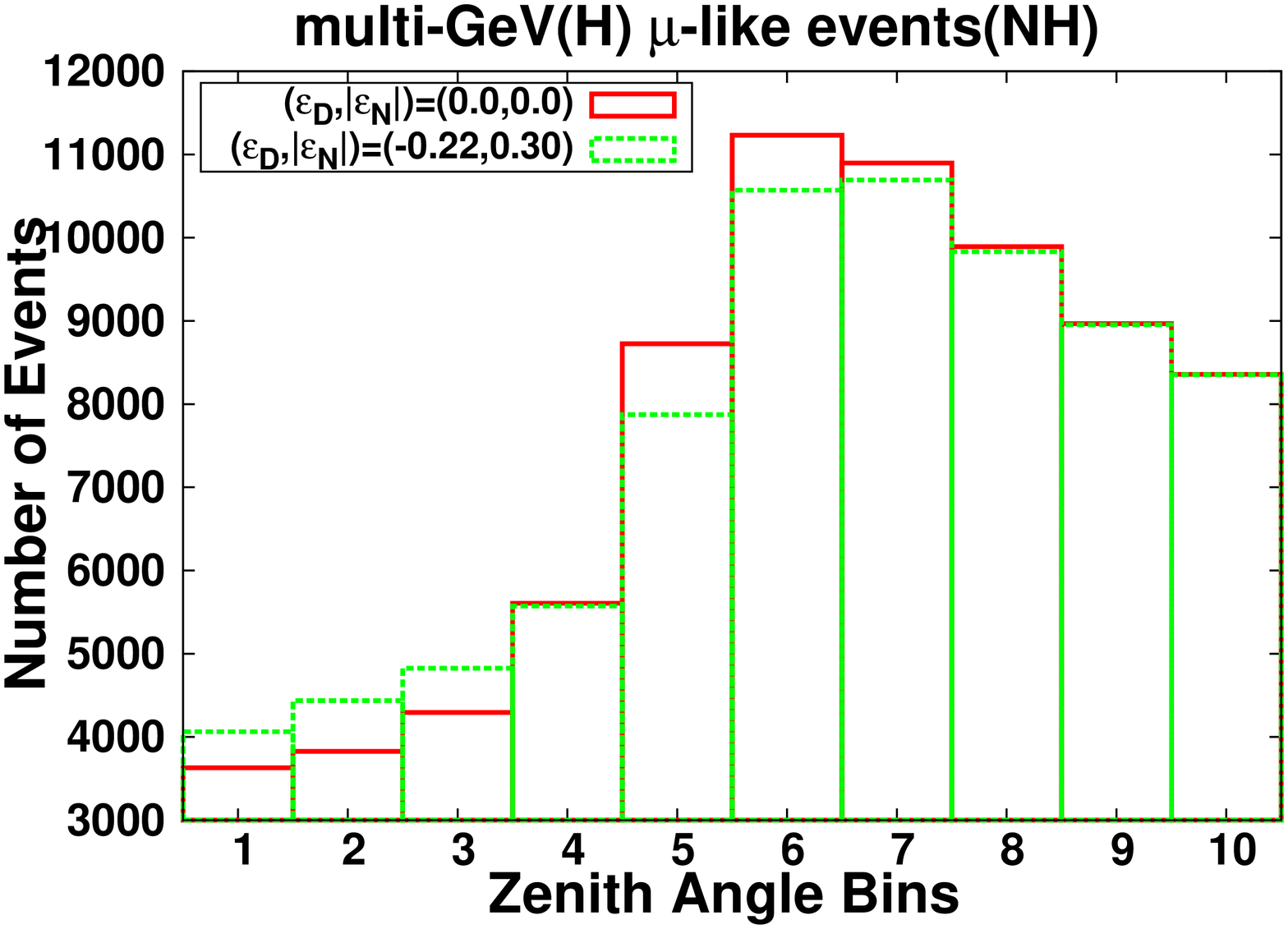}
\includegraphics[scale=0.3,angle=-90]{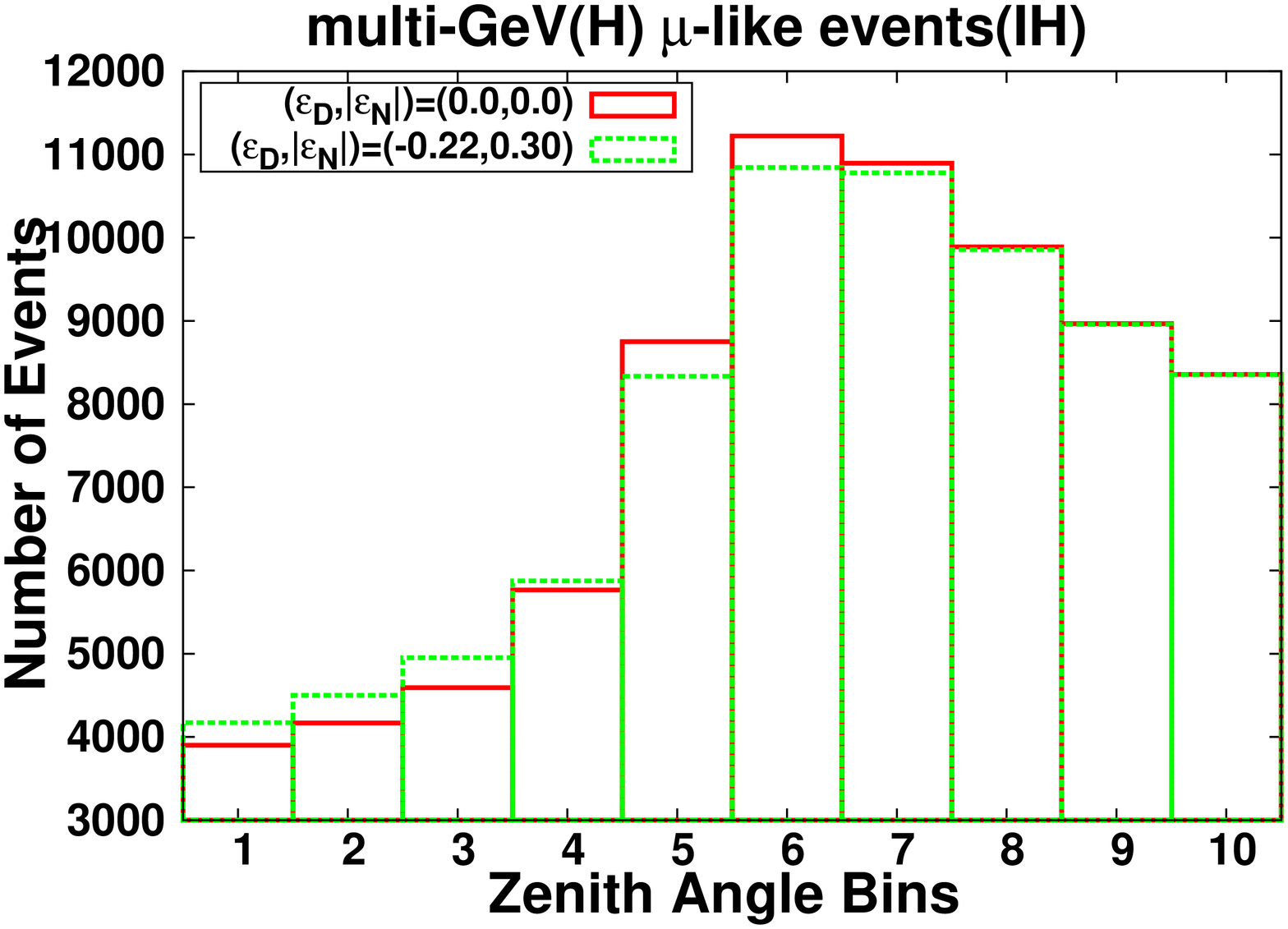}
\caption{The number of the high-energy-bin multi-GeV $\mu$-like events (the red and green boxes are the standard scenario and the scenario with NSI, respectively). The horizontal axis is the zenith angle bin (1 for $-1.0<\cos\Theta<-0.8$, $\dots$, 10 for $0.8<\cos\Theta<1.0$ ). }
\label{fig:fig3}
\end{figure}

To see which bin contributes to $\chi^2$ most, we focused on the number of events difference between the standard scenario and the scenario with NSI (the red and black circle points in Fig.\,\ref{fig:fig1}).
Then we found that the multi-GeV $\mu$-like events coming from the below in the high-energy-bin most contributes to $\chi^2$.
This is because difference between the oscillation probability with NSI and without NSI is large where the neutrino energy is above 10 GeV.
We plotted the numbers of events for the multi-GeV $\mu$-like events in the high-energy-bin in Fig.\,\ref{fig:fig3}. 

\section{Conclusion\label{conclusion}}

In this paper we have studied the sensitivity
of the future HK atmospheric neutrino
experiment to NSI which is suggested by
the tension between the mass squared differences
from the solar neutrino and KamLAND data.
If nature is described by the best fit point
of the combined analysis 
of the solar neutrino and KamLAND data, then
HK will have an affirmative signal at more than
$11\sigma$ ($8\sigma$) in the case of the normal
(inverted) hierarchy, while if she is
represented by the best fit point
of the global analysis, then
HK will have an affirmative signal at
$5.0\sigma$  ($1.4\sigma$) in the case of the normal
(inverted) hierarchy.
We have shown that the channel which is most
sensitive to NSI is the $\mu$-like multi-GeV high energy bin.
This is because the matter effect becomes
most important when the contribution
of the mass squared difference divided by
the neutrino energy becomes comparable
to the matter effect $G_FN_e$.

The present study is the extension of
our previous one\,\cite{Fukasawa:2015jaa}
in the sense that
all the $\epsilon_{\alpha\beta}$ components of the NSI,
including $\epsilon_{e\mu}$, $\epsilon_{\mu\mu}$
and $\epsilon_{\mu\tau}$,
are taken into account and that
$\epsilon_{\tau\tau}$ is not assumed
to be dependent on other components.

In the process of our analysis, we have taken into account
the mapping from the standard parametrization
$\epsilon_{\alpha\beta}~(\alpha, \beta= e, \mu, \tau)$
to the ($\epsilon_D$, $\epsilon_N$)
parameters, which were introduced for the solar neutrino study.
In the simplest approximation in which
$\epsilon_{\alpha\mu}~(\alpha = e, \mu, \tau)$,
$\theta_{13}=0$ and $\theta_{23}=\pi/4$,
it was shown that the allowed region
$|\epsilon_{e\tau}|/|1+\epsilon_{ee}|\lesssim \tan\beta$
in the standard parametrization
from the atmospheric neutrino data
corresponds to the region
$|\epsilon_N|/|1/6-\epsilon_D|\lesssim 
\tan\{2\tan^{-1}(\tan\beta/\sqrt{2})\}$
in the ($\epsilon_D$, $\epsilon_N$) plane.

It is remarkable that
the tension, which was found in
the low energy neutrino data ($E_\nu~\sim$ several MeV)
of the solar and KamLAND experiments,
can be tested by the high energy neutrino data
($E_\nu\sim{\cal O}$(10) GeV)
in the future atmospheric
neutrino experiments with high statistics
through the matter effect.

\appendix

\section{The relation between the standard parametrization
$\epsilon_{\alpha\beta}$ and ($\epsilon_{D}$, $\epsilon_{N}$)}
\label{appendixa}

In this appendix we discuss the relation between the standard parametrization
$\epsilon_{\alpha\beta}$ and the set of the parametrizations
($\epsilon^f_{D}$, $\epsilon^f_{N}$) in the simplest case.
For simplicity we
set $\theta_{13}=0$, $\theta_{23}=\pi/4$,
$\epsilon_{\alpha\mu}=0~(\alpha=e, \mu, \tau)$,
which is a good approximation to some extent.
Then, noting that $\epsilon_{\alpha\beta}=3\epsilon_{\alpha\beta}^d$ and
$\epsilon_{D}=\epsilon_{D}^d$,
$\epsilon_{N}=\epsilon_{N}^d$,
Eq.\,(\ref{epsilonn}) becomes
\begin{eqnarray}
&{\ }&
\hspace*{-60mm}
3\epsilon_{D} = -\frac{1}{2}\epsilon_{e e}
+\frac{1}{4}\epsilon_{\tau \tau}
\label{eq:mapping_fl1}\\
&{\ }&
\hspace*{-60mm}
3\epsilon_{N} = -\frac{1}{\sqrt{2}}\,\epsilon_{e\tau}\,.
\label{eq:mapping_fl2}
\end{eqnarray}
The two nonvanishing eigenvalues of the
matter potential
$\lambda_{e'}$, $\lambda_{\tau'}$ in the
unit of $\sqrt{2}G_FN_e$ are given by
\begin{eqnarray}
&{\ }&
\hspace*{-10mm}
\left(\begin{array}{c}
\lambda_{e'}\cr
\lambda_{\tau'}
\end{array}\right)=
\frac{1+\epsilon_{ee}+\epsilon_{\tau\tau}}{2}\pm
\sqrt{\left(\frac{1+\epsilon_{ee}-\epsilon_{\tau\tau}}{2}\right)^2
+|\epsilon_{e\tau}|^2}\,,
\nonumber
\end{eqnarray}
and they satisfy the following relations:
\begin{eqnarray}
&{\ }&
\hspace*{-50mm}
\lambda_{e'}+\lambda_{\tau'}=
1+\epsilon_{ee}+\epsilon_{\tau\tau}
\label{root1}\\
&{\ }&
\hspace*{-50mm}
\lambda_{e'}\,\lambda_{\tau'}=
(1+\epsilon_{ee})\,\epsilon_{\tau\tau}-|\epsilon_{e\tau}|^2
\label{root2}
\end{eqnarray}
Assuming $1+\epsilon_{ee}>0$, $\epsilon_{\tau\tau}>0$,
we postulate the following approximate parabolic relation:
\begin{eqnarray}
&{\ }&
\hspace*{-30mm}
\lambda_{\tau'}=
\frac{1+\epsilon_{ee}+\epsilon_{\tau\tau}}{2}-
\sqrt{\left(\frac{1+\epsilon_{ee}-\epsilon_{\tau\tau}}{2}\right)^2
+|\epsilon_{e\tau}|^2}=\alpha~(>0)\,.
\label{quasiparabolic}
\end{eqnarray}
From Eqs.\,(\ref{root1}) and (\ref{root2}) we have
\begin{eqnarray}
&{\ }&
\hspace*{-20mm}
\lambda_{e'}=1+\epsilon_{ee}+\epsilon_{\tau\tau}-\alpha
\nonumber\\
&{\ }&
\hspace*{-14mm}
=\frac{(1+\epsilon_{ee})\,\epsilon_{\tau\tau}-|\epsilon_{e\tau}|^2}
{\alpha}
=\frac{(1+\epsilon_{ee})\,\epsilon_{\tau\tau}-|3\sqrt{2}\epsilon_N|^2}
{\alpha}\,.
\label{eq:mapping_fl3}
\end{eqnarray}
From Eq.\,(\ref{eq:mapping_fl1}) we obtain
\begin{eqnarray}
&{\ }&
\hspace*{-60mm}
1+\epsilon_{ee}=\left(1-6\epsilon_{D}\right)
+\frac{1}{2}\epsilon_{\tau\tau}\,.
\label{eq:mapping_fl4}
\end{eqnarray}
Substituting Eq.\,(\ref{eq:mapping_fl4})
into Eq.\,(\ref{eq:mapping_fl3}), we get
\begin{eqnarray}
&{\ }&
\hspace*{-10mm}
\frac{1}{\alpha}
\left\{\left(1-6\epsilon_{D}
+\frac{1}{2}\epsilon_{\tau\tau}\right)\epsilon_{\tau\tau}
-|3\sqrt{2}\epsilon_N|^2
\right\}
=1-6\epsilon_{D}
+\frac{3}{2}\epsilon_{\tau\tau}-\alpha\,,
\nonumber
\end{eqnarray}
which yields
\begin{eqnarray}
&{\ }&
\hspace*{-1mm}
\epsilon_{\tau\tau}-\alpha
=-\left(1-6\epsilon_{D}\right)
+\frac{\alpha}{2}
+\left\{\left(1-6\epsilon_{D}
-\frac{\alpha}{2}\right)^2
+4|3\epsilon_N|^2
\right\}^{1/2}
\nonumber\\
&{\ }&
\hspace*{-1mm}
1+\epsilon_{ee}-\alpha=
\frac{1}{2}\left(1-6\epsilon_{D}\right)-\frac{\alpha}{4}
+\frac{1}{2}\left\{\left(1-6\epsilon_{D}
-\frac{\alpha}{2}\right)^2
+4|3\epsilon_N|^2
\right\}^{1/2}\,.
\nonumber
\end{eqnarray}
It is easy to see that the last two equations satisfy
\begin{eqnarray}
&{\ }&
\hspace*{-60mm}
\left(1+\epsilon_{ee}-\alpha\right)
\left(\epsilon_{\tau\tau}-\alpha\right)
=2|3\epsilon_N|^2\,.
\label{eq:mapping_fl5}
\end{eqnarray}
Eq.\,(\ref{eq:mapping_fl5}), which is
the eigenvalue equation for the two nonvanishing eigenvalues
of the matter matrix,
should be
satisfied because $\lambda_{\tau'}=\alpha$
is one of the two nonvanishing eigenvalues.
Eq.\,(\ref{eq:mapping_fl5}) can be regarded as
the generalized parabolic relation in the case
of nonvanishing $\alpha$, and it reduces
to Eq.\,(\ref{eq:ansatz_a}) in the limit
$\alpha\to 0$.

Eq.(\ref{eq:mapping_fl5}) suggests
that the matter angle in the case of
nonvanishing $\alpha$ should be defined as
\begin{eqnarray}
&{\ }&
\hspace*{-10mm}
\tan\beta=
\frac{\left|\epsilon_{e\tau}\right|}
{1+\epsilon_{ee}-\alpha}
\nonumber\\
&{\ }&
\hspace*{-0mm}
=\frac{\left|3\sqrt{2}\epsilon_N\right|}
{
1/2-3\epsilon_{D}-\alpha/4
+\left\{\left(1/2-3\epsilon_{D}
-\alpha/4\right)^2
+|3\epsilon_N|^2
\right\}^{1/2}
}\,.
\nonumber
\end{eqnarray}
Here if we introduce a new angle
\begin{eqnarray}
&{\ }&
\hspace*{-90mm}
\tan\beta'\equiv\frac{\tan\beta}{\sqrt{2}}\,,
\nonumber
\end{eqnarray}
then 
\begin{eqnarray}
&{\ }&
\hspace*{-0mm}
\tan\beta'=
\frac{\left|3\epsilon_N\right|}
{
1/2-3\epsilon_{D}-\alpha/4
+\left\{\left(1/2-3\epsilon_{D}
-\alpha/4\right)^2
+|3\epsilon_N|^2
\right\}^{1/2}
}
\nonumber\\
&{\ }&
\hspace*{10mm}
=\frac{
-\left(1/2-3\epsilon_{D}-\alpha/4\right)
+\left\{\left(1/2-3\epsilon_{D}
-\alpha/4\right)^2
+|3\epsilon_N|^2
\right\}^{1/2}}
{\left|3\epsilon_N\right|}
\,.
\label{eq:mapping_fl6}
\end{eqnarray}
From Eq.\,(\ref{eq:mapping_fl6}) we have
\begin{eqnarray}
&{\ }&
\hspace*{-10mm}
\tan2\beta'=
\frac{2\tan\beta'}
{1-\tan^2\beta'}
\nonumber\\
&{\ }&
\hspace*{4mm}
=\frac{2\left|3\epsilon_N\right|\left\{
\displaystyle\sqrt{\left(\frac{1}{2}-3\epsilon_{D}
-\displaystyle\frac{\alpha}{4}\right)^2
+|3\epsilon_N|^2}-\left(\frac{1}{2}-3\epsilon_{D}
-\frac{\alpha}{4}\right)\right\}
}
{|3\epsilon_N|^2-\left\{
\displaystyle\sqrt{\left(\frac{1}{2}-3\epsilon_{D}
-\frac{\alpha}{4}\right)^2
+|3\epsilon_N|^2}
-\left(\frac{1}{2}-3\epsilon_{D}
-\frac{\alpha}{4}\right)\right\}^2}
\nonumber\\
&{\ }&
\hspace*{4mm}
=\frac{\left|3\epsilon_N\right|}
{\displaystyle\frac{1}{2}-3\epsilon_{D}
-\frac{\alpha}{4}}\,.
\label{eq:mapping_fl7}
\end{eqnarray}
Eq.\,(\ref{eq:mapping_fl7}) implies that
the allowed region of the atmospheric
neutrino experiment with
the generalized parabolic relation
(\ref{eq:mapping_fl5}) is the one
surrounded by the $\epsilon_N=0$ axis
and the straight line 
$|\epsilon_N|=|\tan2\beta'||1/2-3\epsilon_{D}-\alpha/4|$
with the gradient $|\tan2\beta'|$ and 
the $x$-intercept $\epsilon_D=1/6-\alpha/12$.
In the limit $\alpha\to 0$, Eq.\,(\ref{eq:mapping_fl7})
reduces to Eq.\,(\ref{gradbetaprime}).

\section*{Acknowledgments}
This research was partly supported by a Grant-in-Aid for Scientific
Research of the Ministry of Education, Science and Culture, under
Grants No. 24540281 and No. 25105009.

\end{document}